\documentclass{PoS}
\usepackage[utf8]{inputenc}
\usepackage{amsmath,amsfonts,amssymb,mathrsfs,fixmath,amsthm}
\usepackage{bbm}

\usepackage{graphics}
\graphicspath{{./plots/}}
\usepackage{subcaption}

\usepackage[sort&compress]{natbib}
\bibpunct{[}{]}{,}{n}{}{}
\makeatletter

\DeclareCaptionLabelFormat{mysublabelfmt}{(\alph{sub\@captype})}
\makeatother

\newcommand{\Dov}{\ensuremath{D_\text{ov}}}
\newcommand{\Dw}{\ensuremath{D_\text{w}}}
\newcommand{\Db}{\ensuremath{\overline{D}}}
\newcommand{\K}{\ensuremath{\mathcal{K}}}
\newcommand{\ket}[1]{\ensuremath{\left|#1\right\rangle}}
\newcommand{\bra}[1]{\ensuremath{\left\langle#1\right|}}
\newcommand{\braket}[2]{\ensuremath{\left\langle\left.#1\right|  #2 \right\rangle}}
\newcommand{\norm}[1]{\ensuremath{\sqrt{\braket{#1}{#1}}}}
\newcommand{\+}[1]{\ensuremath{\ket{#1}}}
\DeclareMathOperator{\sgn}{sgn}
\DeclareMathOperator{\diag}{diag}

\newtheorem{theorem}{Theorem}

\title{A Method to Calculate Conserved Currents and Fermionic Force for the Lanczos Approximation to the Overlap Dirac Operator}

\ShortTitle{Taking derivatives of the overlap operator}

\author{\speaker{Matthias Puhr}\\
        Institut für Theoretische Physik, Universität Regensburg \\
        E-mail: \email{matthias.puhr@physik.uni-regensburg.de}}

\author{Pavel Buividovich\\
       Institut für Theoretische Physik, Universität Regensburg\\
       E-mail: \email{pavel.buividovich@physik.uni-regensburg.de}}

\abstract{The overlap Dirac operator obeys the Ginsparg-Wilson equation and offers a possibility to introduce chiral symmetry on the lattice. Evaluating the overlap operator is numerically very expensive and one has to rely on approximation methods. At finite chemical potential the overlap operator can be efficiently computed with the two-sided Lanczos algorithm. 
To calculate conserved currents on the lattice, or to evaluate the fermionic force in HMC calculations, one needs to compute derivatives of the Dirac operator with respect to gauge fields. 
In this paper we present a method to simultaneously compute the action of the overlap operator and its derivative on a source vector. }

\FullConference{The 32nd International Symposium on Lattice Field Theory,\\
		23-28 June, 2014\\
		Columbia University New York, NY}

\begin{document}

\section{Motivation}
\label{sec:motivation}
Quantum chromodynamics (QCD) is a theory that is chirally symmetric in the limit of massless quarks. Chiral symmetry and its spontaneous breaking play an important role in QCD phenomenology. Many observables depend strongly on the chiral properties of QCD. When one studies QCD on a finite space-time lattice it is therefore desirable to use a discretisation of the Dirac operator that respects chiral symmetry and is free of doublers.
A chiral symmetry preserving and doubler free discretisation of the Dirac operator has to obey the Ginsparg-Wilson equation \cite{Ginsparg1982}. Finding such a lattice Dirac operator is a non trivial task, but today several solutions are known. One of them is the overlap Dirac operator, which is an exact solution to the Ginsparg-Wilson equation. Unfortunately the definition of the overlap operator includes the matrix sign function, which is numerically very expensive. Evaluating the overlap operator exactly is therefore not feasible for reasonably large lattice sizes and one has to rely on approximation methods. 

A very efficient method is the two-sided Lanczos (TSL) algorithm. It computes an approximation to the action of the overlap operator on a source vector. A major advantage of the TSL algorithm is that it works for general complex matrices. This is important when one considers finite chemical potential, where the sign function of a non-Hermitian matrix has to be evaluated\cite{Bloch2006}. 
While the TSL method is well suited to compute the overlap operator, the evaluation of the fermionic force in HMC calculations and the calculation of conserved currents make it necessary to additionally compute the derivative of the Dirac operator.
The purpose of this paper is to introduce a numerical method to simultaneously compute the action of the overlap Dirac operator and its derivative on a source vector. 

\section{The overlap operator and the matrix sign function}
\label{sec:overlap}
 At finite quark chemical potential $\mu$ the massless overlap Dirac operator is defined as\cite{Bloch2006}
\begin{equation}
   \label{eq:overlap}
   \Dov := \frac{1}{a}\left(\mathbbm{1}+\gamma_5 \sgn\left[ \gamma_5 \Dw(\mu) \right] \right),
\end{equation}
 where $\Dw(\mu)$ is the Wilson Dirac operator at non-zero chemical potential, $\sgn$ stands for the matrix sign function and $a$ is the lattice spacing. The function $f$ of a matrix $A \in \mathbb{C}^{n\times n}$ can be defined in several equivalent ways \cite{Higham2008}. For a diagonalisable\footnotemark \ $A$, that is $A=U\Lambda U^{-1}$ with the diagonal eigenvalue matrix $\Lambda = \diag(\lambda_1,\cdots,\lambda_n)$, one can employ the particularly simple and convenient spectral decomposition:
  \begin{equation}
  \label{eq:spectral}
  f(A) :=U f(\Lambda)U^{-1}  ~ , \quad f(\Lambda):=diag(f(\lambda_1),\cdots,f(\lambda_n))
\end{equation}
\footnotetext{This can be generalised to non-diagonalisable matrices using the Jordan canonical form\cite{Higham2008}.}
In general the argument of the sign function $\gamma_5\Dw(\mu)$ is a non-Hermitian matrix with complex eigenvalues $\lambda_i$ and we need a generalisation of the sign function for complex arguments. It is important that  the sign function satisfies $\sgn(z)^2=1$ for any complex number $z$, since this ensures that the overlap operator respects the Ginsparg-Wilson relation. Moreover for $x \in \mathbb{R}\setminus \{0\}$ the complex sign function should reduce to the standard definition $\sgn(x)=\pm 1$. A choice that has the requested properties is 
\begin{equation}
  \label{eq:sgn}
  \sgn(z):=\frac{z}{\sqrt{z^2}}=\sgn(\Re(z)),
\end{equation}
where the cut of the square root is chosen along the negative real axis, so that the cut of the sign function is along the imaginary axis.

For numerical matrix computations it is often more convenient to use an iterative method to evaluate the sign function, the so called Roberts iteration:
\begin{equation}
  \label{eq:roberts}
  X_{k+1}:=\frac{1}{2}\left(X_k + X_k^{-1}\right) \ , \quad X_0 := A
\end{equation}
This is Newton's method applied to the matrix equation $X^2=1$. If $A$ has no purely imaginary eigenvalues the $X_k$ converge quadratically to $\sgn(A)$\cite{Higham2008}.

The run-time complexity of both the spectral decomposition \eqref{eq:spectral} and the Roberts iteration \eqref{eq:roberts} is approximately $\mathcal{O}(n^3)$. Therefore the numerical cost of an evaluation of the overlap operator becomes prohibitively large very quickly. Even for relatively small lattice sizes it is not feasible to compute the matrix sign function and one has to resort to approximation methods.  

\section{The two-sided Lanczos algorithm}
In many applications it is not necessary to compute a matrix function $f(A)$ explicitly since it is sufficient to evaluate the action of $f(A)$ on a vector $\+\psi$. The TSL algorithm is a Krylov subspace method and computes an approximation to $\+y=f(A)\+\psi$. The Krylov subspace of order $k$ for a matrix $A$ and a vector $\+\psi$ is defined as 
\begin{equation}
  \label{eq:krylov}
  \K_k(A,\+\psi):=\operatorname{span}(\+\psi,A\+\psi,\cdots,A^{k-1}\+\psi).
\end{equation}
The TSL constructs biorthonormal bases $V_k=(\+{v_1},\cdots,\+{v_k})$ and $W_k=(\+{w_1},\cdots,\+{w_k})$ of the right (\,$\K_k(A,\+\psi)$\,) and left (\,$\K_k(A^{\dagger},\+\psi)$\,) Krylov subspace, such that 
\begin{equation}
  \label{eq:td1}
  T_k := W_k^\dagger A V_k  
\end{equation}
is a tridiagonal $k \times k $ matrix:
\begin{equation}
  \label{eq:td2}
  T_k=
  \begin{pmatrix}
    \alpha_1 & \gamma_1 & 0      & \dots   & 0           \\
    \beta_1  & \alpha_2 & \ddots & \ddots & \vdots       \\
    0        & \ddots   & \ddots & \ddots & 0            \\
    \vdots   & \ddots   & \ddots & \ddots & \gamma_{k-1} \\
    0        & \dots    & 0      & \beta_{k-1} & \alpha_k 
  \end{pmatrix}
\end{equation}
Using equations \eqref{eq:td1} and \eqref{eq:td2} one can show  that $V_k$ and $W_k$ can be built with the following recurrence relations \cite{Golub1996} :
\begin{equation}
  \label{eq:3}
\begin{aligned}
   \beta_i\+{v_{i+1}}       & =   (A - \alpha_i)\+{v_i} - \gamma_{i-1}\+{v_{i-1}} \\ 
   \gamma_i^* \+{w_{i+1}}   & =  (A^\dagger - \alpha_i^*)\+{w_i} - \beta_{i-1}^* \+{w_{i-1}}
\end{aligned}
\end{equation}
The seeds of the recurrence relations (\ref{eq:3}) are chosen to be $\+{v_1}=\+{w_1}= \+\psi/\|\psi\|$, with the norm $\|\psi\|:=\norm{\psi}$. The diagonal of $T_k$ is fixed by $\alpha_i = \bra{w_i} A \+{v_i}$, whereas $\beta_i$ and $\gamma_i$ are not uniquely determined  and can be obtained from the normalisation condition $\braket{w_{i+1}}{v_{i+1}}=1$.

The matrix $V_kW_k^\dagger$ is an oblique projector on the space $\K_k(A,\+\psi)$ and in a first step we approximate $\+y$ by the oblique projection of $f(A)\+\psi$ on $\K_k(A,\+\psi)$:
\begin{equation}
  \label{eq:oblique_pro}
   \+y \approx \+y_{\text{obl}} = V_kW_k^\dagger  f(A) V_kW_k^\dagger \+\psi.
\end{equation}
Combined with the approximation $W_k^\dagger  f(A) V_k \approx f(T_k)$  this yields the final result\footnotemark{} 
\begin{equation}
  \label{eq:approx}
  \+y \approx \|\psi\| V_kf(T_k) \+{e_1}.
\end{equation}
\footnotetext{Note that by construction $\braket{w_i}{\psi} = \|\psi\|\+{e_1} \delta_{i1}$}
The problem of evaluating $f(A)$ is now reduced to the calculation of the bases $V_k$ and $W_k$ and the computation of $f(T_k)$ for the tridiagonal matrix $T_k$. The complexity of the function evaluation is reduced from $\mathcal{O}(n^3)$ to $\mathcal{O}(nk) + \mathcal{O}(k^3)$. In practice one can obtain very good approximations of the overlap Dirac operator already for $k \ll n$. The efficiency of the TSL can be further increased by using deflation methods \cite{Bloch2007,Bloch2008} and a nested version of the algorithm \cite{Bloch2011}. 

\section{Derivatives of the Lanczos Algorithm}
\subsection{The method}
\begin{figure}[h]
  \centering
  \begin{subfigure}[h]{0.49\textwidth}
    \includegraphics[width=\textwidth]{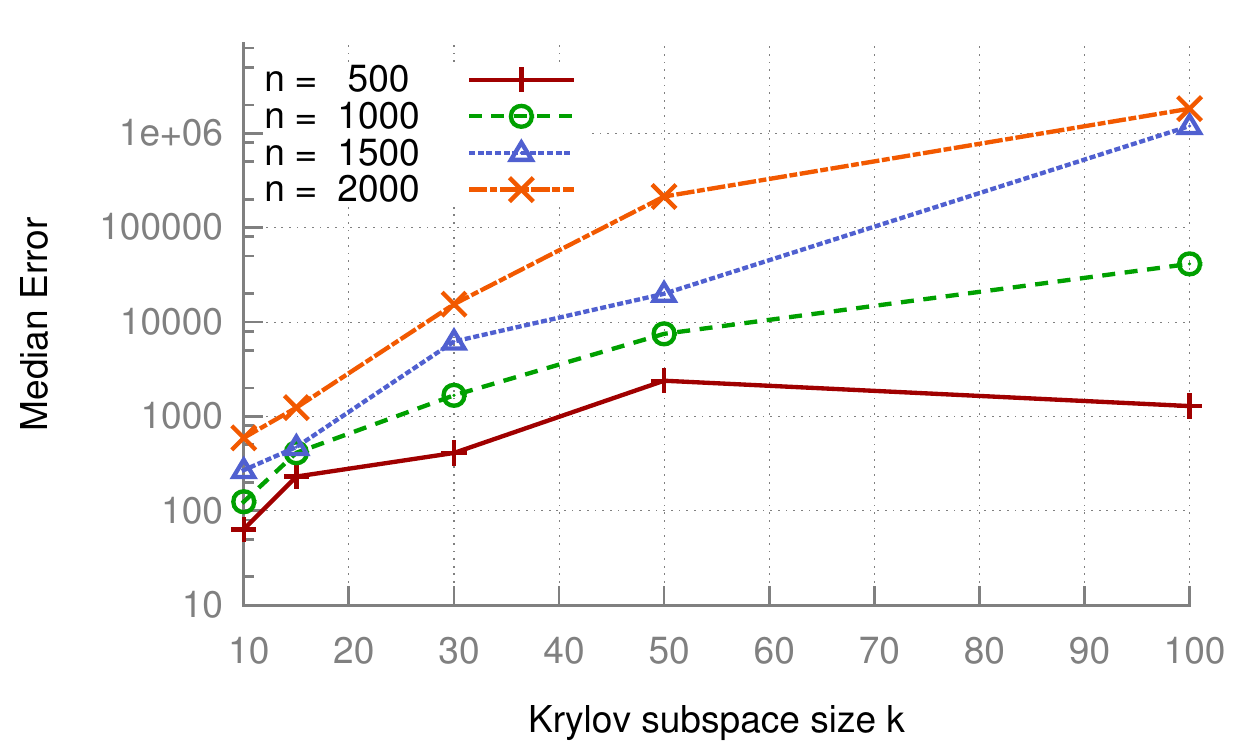}
    \caption{Algorithmic differentiation error }
    \label{fig:alg_diff}
  \end{subfigure}
  \begin{subfigure}[h]{0.49\textwidth}
    \includegraphics[width=\textwidth]{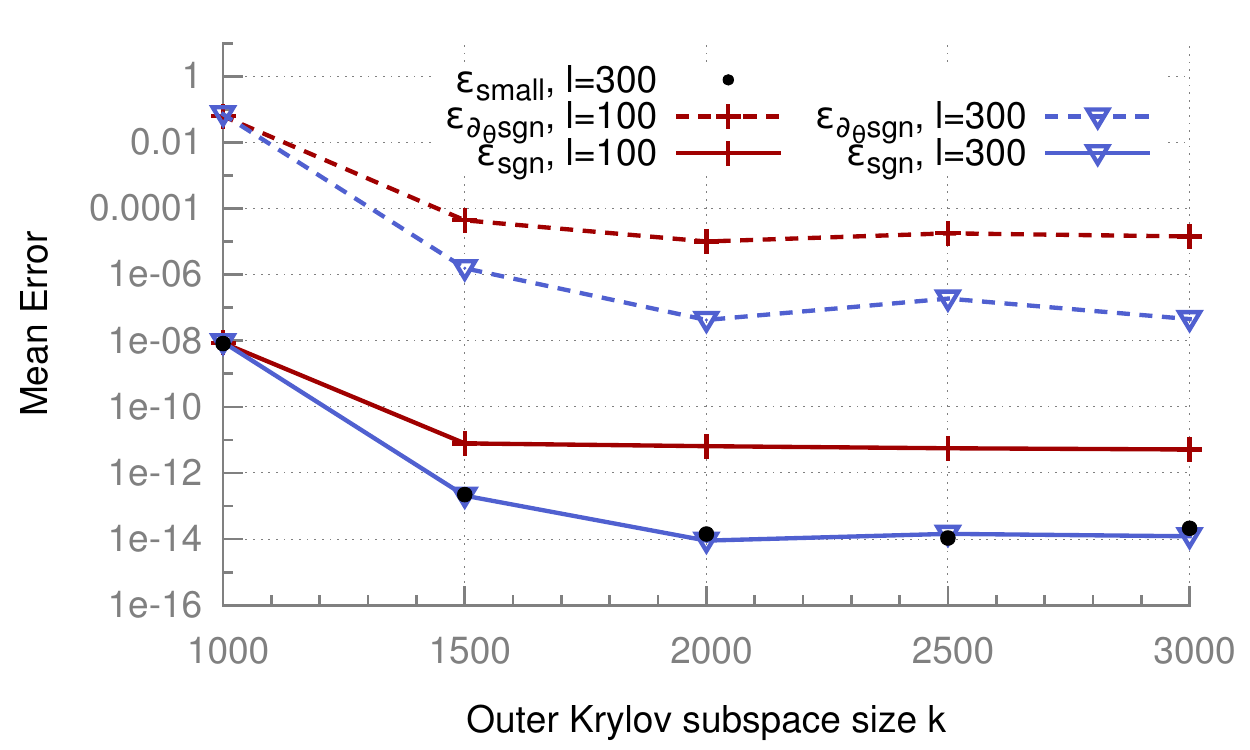}
    \caption{$6\text{x}6^3$ Lattice}
    \label{fig:6x6}
  \end{subfigure}
  \begin{subfigure}[h]{0.49\textwidth}
    \includegraphics[width=\textwidth]{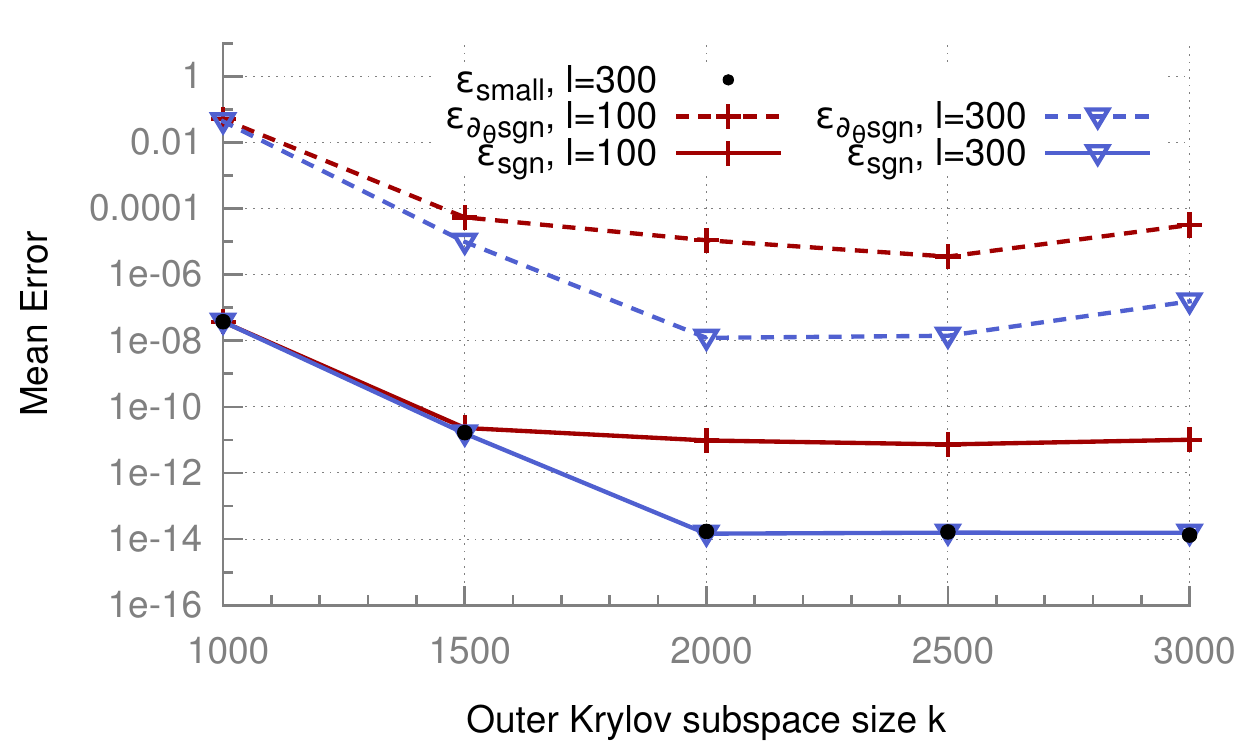}
    \caption{$8\text{x}8^3$ Lattice}
    \label{fig:8x8}
  \end{subfigure}
\begin{subfigure}[h]{0.49\textwidth}
    \includegraphics[width=\textwidth]{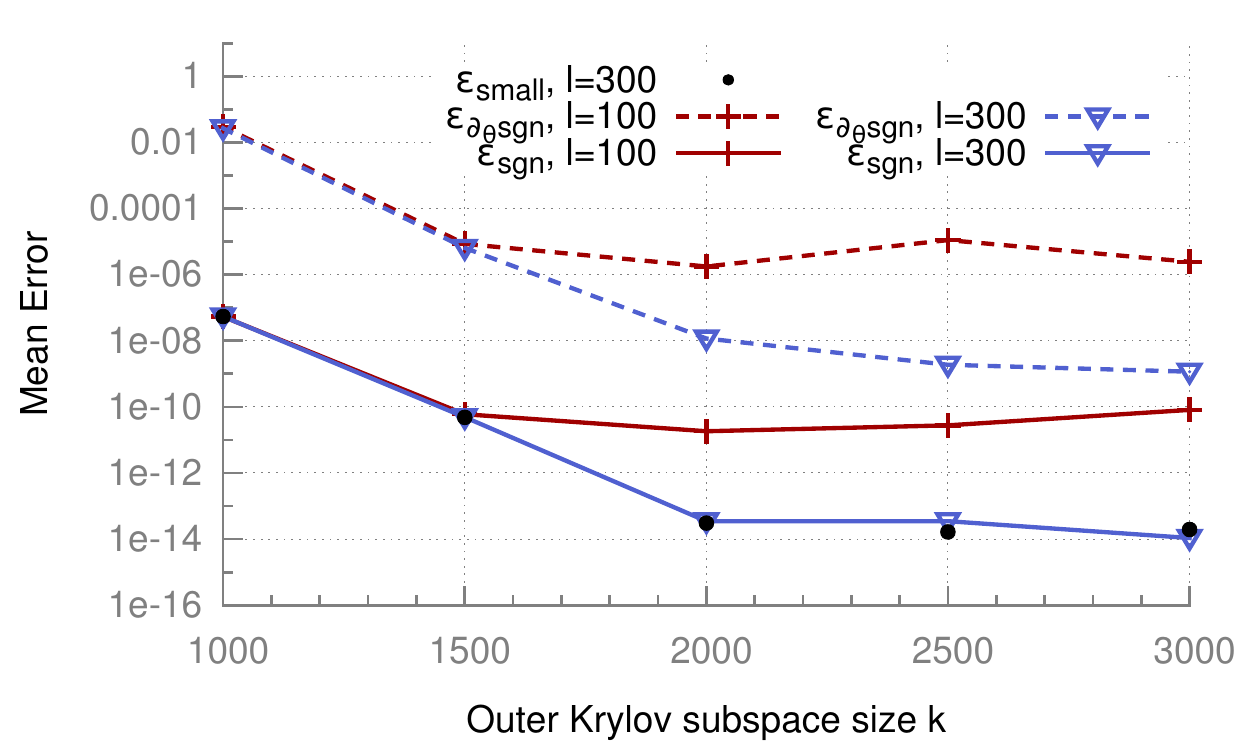}
    \caption{$12\text{x}12^3$ Lattice}
    \label{fig:12x12}
  \end{subfigure}
  \caption[]{ \subref{fig:alg_diff} Algorithmic differentiation of the TSL method for $n$-dimensional test matrices. The algorithmic differentiation approach is numerically unstable and the errors increase with the Krylov subspace size. \subref{fig:6x6} -- \subref{fig:12x12} Results for $\sgn(\gamma_5 \Dw)$  and its derivative at finite chemical potential $\mu=0.20$ (in lattice units). We plot the errors (see section \ref{sec:results})  as a function of the outer Krylov subspace size for two different inner Krylov subspace sizes $l$. The points are connected to guide the eye. } 
\label{figure}
\end{figure}

It is relatively simple to compute the derivatives $\partial \Dw / \partial \theta_\nu(x)$ of the Wilson Dirac operator over the (lattice) gauge field $\theta_\nu(x)$ by hand. For the overlap operator things are more complicated and if approximation methods like TSL are used the derivatives can only be evaluated numerically. Apart from divided difference methods the most straightforward way to calculate derivatives of the TSL is algorithmic differentiation\cite{Griewank2008}. However, experiments with random matrices showed that the algorithmic differentiation approach for TSL is numerically unstable, see figure \ref{fig:alg_diff}. The reason for this is most probably the loss of biorthogonality of $V_k$ and $W_k$ because of round-off errors. While this is not a big problem for the TSL, it seems to strongly influence the numerical stability of the algorithmic differentiation. 

To avoid the numerical stability problems we propose an algorithm based on the following matrix function theorem\cite{Mathias1996}:
\begin{theorem}
\label{th:mathias}
   Let $A(t) \in \mathbb{C}^{n\times n}$ be differentiable at $t=0$ and assume that the spectrum of $A(t)$ is contained in  an open subset $\mathcal{D}\subset\mathbb{C}$ for all $t$ in some neighbourhood of $0$. Let $f$  be $2n-1$ times continuously differentiable on~$\mathcal{D}$. Then:
    $$ f\left(\bar{A}\right) \equiv \left[\begin{array}{cc}
        f(A(0)) & \left. \frac{d}{dt} \right|_{t=0} f(A(t)) \\
        0 & f(A(0))
      \end{array}\right], \quad \bar{A} := \left[
        \begin{array}{cc}
          A(0) & {\dot{A}}(0) \\
          0   & A(0) 
        \end{array}
      \right] $$
\end{theorem}
\noindent Theorem \ref{th:mathias} relates the derivative of a matrix function to the function of a  block matrix. It is remarkable that this enables us to compute the derivative of $f$ without the knowledge of $f'$. This comes at the cost of evaluating the function for a matrix that has twice the dimension of the original matrix \nolinebreak[1] $A$. Fortunately the block matrix $\bar{A}$ is sparse and one only needs to store $A$ and $\dot{A}$, as in the algorithmic differentiation approach. Another concern is the convergence of the Lanczos algorithm for the block matrix $\bar{A}$. Convergence critically depends on the spectrum of the matrix. It is easy to proof that $\bar{A}$ has the same spectrum as the matrix $A$ and every eigenvalue of $A$ is a (degenerate) eigenvalue of $\bar{A}$. In general matrices of the special form $\bar{A}$ are not diagonalisable. Numerical experiments show that the Jordan normal form of $\bar{A}$ has non-trivial Jordan blocks of size exactly two, i.e. the Jordan matrix is a block diagonal matrix with 2-dimensional matrices on the main diagonal and zeros everywhere else.   
Note that the formula for the eigenvectors of $\bar{A}$ presented in the talk was derived under the assumption of diagonalisability and does not hold in the general case.

The complex sign function is differentiable on $\mathbb{C}\setminus i \mathbb{R}$ and in practice $\gamma_5 \Dw$ does not have purely imaginary eigenvalues. Therefore theorem \ref{th:mathias} holds and we have
\begin{equation}
  \label{eq:mathias_sgn}
  \sgn\left(\gamma_5 \Db \right)\left(
  \begin{array}{c}  0 \\ \+\psi  \end{array}\right) = \left(
  \begin{array}{c}
    \frac{\partial }{\partial \theta_\nu(x)}\sgn(\gamma_5 \Dw)\+\psi \\ \sgn(\gamma_5 \Dw)\+\psi
  \end{array}
\right),  \quad \Db := \left[\begin{array}{cc}  \Dw & {\frac{\partial \Dw}{\partial \theta_\nu(x)}} \\  0   & \Dw \end{array} \right].
\end{equation}
We can now apply the TSL method to compute the result of the action of $\sgn(\gamma_5 \Db)$ on the vector $(0,\+\psi)^T$ and to simultaneously obtain an approximation for $\sgn(\gamma_5 \Dw)\+\psi$ and $\frac{\partial }{\partial \theta_\nu(x)}\sgn(\gamma_5 \Dw)\+\psi$. In this way the derivative of the matrix sign function can be computed without the need to modify the TSL algorithm.

\subsection{Numerical Results}
\label{sec:results}

To test the method proposed in the last subsection we compute $\sgn(\gamma_5 \Db)(0,\+\psi)^T$  for random source vectors $\+\psi$ to get an approximation for $(\sgn(\gamma_5 \Dw)\+\psi,\partial_{ \theta_\nu}\sgn(\gamma_5 \Dw)\+\psi)^T$. The lattice site $x$ and the direction $\nu$ for the derivative $\partial / \partial \theta_\nu(x)$ are chosen at random. For the tests we use $SU(3)$ configurations generated with an improved action\cite{Luescher1985} on lattices of small to medium size. The inverse coupling is set to $\beta=5.95$ in all cases. 

To estimate the numerical error $\varepsilon_{\sgn}$ of the sign function approximation we use the identity $\sgn( A)^2  = \mathbbm{1}$ and define $\varepsilon_{\sgn}:= \|\sgn(A)^2\+\psi -\+\psi \|/(2 \|\psi\|)$. The factor two enters the definition because we have to apply our approximation twice to compute the square of the sign function. The commutator $\left\{\partial_{ \theta_\nu}\sgn(A),\sgn(A)\right\}$  vanishes, as can be seen most easily by taking the derivative of the squared sign function. We use this fact to define the numerical error  $\varepsilon_{\partial_\theta\sgn}$ of the derivative as  $\varepsilon_{\partial_\theta\sgn}:= \|\left\{\partial_{ \theta_\nu}\sgn(A),\sgn(A)\right\}\+\psi\|/(2\|\psi\|)$. To compare the convergence properties, we also computed the numerical error $\varepsilon_\text{small}$ of a TSL approximation to $\sgn(\gamma_5 \Dw)\+\psi$.

In our calculations we use a nested version of the TSL method\cite{Bloch2011} with a single nesting step. The outer Krylov subspace size $k$ varies between $k=1000$ and $k=3000$. For the inner subspace size $l$ values between $l=100$ and $l=500$ are used. We find that for our test cases an inner size $l=300$ is sufficient and further increasing the size of the inner space does not significantly improve the approximation.  The results for different lattice sizes are plotted in figures \ref{fig:6x6} -- \ref{fig:12x12}. We observe that $\varepsilon_{\text{small}}$ and $\varepsilon_{\sgn}$ lie almost on top of each other. Although the matrix dimension increases by a factor of two, it is still sufficient to use the same Krylov subspace size to reach a given precision. The algorithm scales very well with matrix size. To reach a precision of $10^{-12}$ in the sign function approximation we need a Krylov subspace size of $k\approx 1400$, $k \approx 1600$ and $k \approx 1800$ for $6\text{x}6^3$, $8\text{x}8^3$ and $12\text{x}12^3$ lattices, respectively.

In all test cases $\varepsilon_{\sgn}$  is much smaller that $\varepsilon_{\partial_\theta\sgn}$, but they stem from different error definitions and are not directly comparable. Qualitatively $\varepsilon_{\sgn}$ and $\varepsilon_{\partial_\theta\sgn}$ show the same behaviour: The error decreases with increasing Krylov subspace size until the optimal subspace size is reached. Then the approximation converges and an additional increase of the subspace size does not improve the results.

\section{Conclusion and Outlook}
\label{sec:conclusion}
We introduce a new numerical method to simultaneously compute the action of a matrix function and its derivative on a source vector and present test results for $\sgn(\gamma_5 \Dw)$ and $\partial_{ \theta_\nu}\sgn(\gamma_5 \Dw)$ on small and medium sized lattices. The method is based on the matrix function identity (\ref{eq:mathias_sgn}) and uses the well established two-sided Lanczos algorithm to compute the action of a matrix function on a vector. Our tests show that the proposed method is reliable and scales very well.

The convergence properties of the TSL for a given matrix $A$ are closely related to the spectrum of $A$. 
The efficiency of the algorithm can be greatly enhanced by using deflation methods\cite{Bloch2007,Bloch2008}. 
We are currently working on the implementation of suitable deflation techniques for the proposed method. One problem is that the block matrix constructed as part of the algorithm is in general not diagonalisable, which makes the adaption of standard deflation methods difficult. A version of theorem \ref{th:mathias} for higher order derivatives exists\cite{Mathias1996} and we are investigating ways to generalise our method to higher order derivatives.

\section*{Acknowledgements}
\label{sec:ack}
We acknowledge helpful discussions with Andreas Frommer, who pointed us to theorem \ref{th:mathias}, Jacques Bloch and Simon Heybrock. We thank Oleg Kochetkov for providing the gauge configurations.

\end{document}